# Anomalous polarization dependence of Raman scattering and crystallographic orientation of black phosphorus

*Jungcheol Kim,* [a, ‡] *Jae-Ung Lee,* [a, ‡] *Jinhwan Lee,* [b, ‡] *Hyo Ju Park,*[c] *Zonghoon Lee,*[c] *Changgu Lee* [b, d ,*] *and Hyeonsik Cheong* [a,*]

[a] Department of Physics, Sogang University, Seoul 121-742, Korea

[b] Department of Mechanical Engineering and Center for Human Interface Nano Technology (HINT), Sungkyunkwan University, Suwon, 440-746, Korea

[c] School of Materials Science and Engineering, Ulsan National Institute of Science and Technology (UNIST), Ulsan 689-798, Korea

[d] SKKU Advanced Institute of Nanotechnology (SAINT), Sungkyunkwan University, Suwon, 440-746, Korea

[*]E-mail: hcheong@sogang.ac.kr and peterlee@skku.edu

**ABSTRACT**

We investigated polarization dependence of the Raman modes in black phosphorus (BP) using five different excitation wavelengths. The crystallographic orientation was determined by comparing polarized optical microscopy with high-resolution transmission electron microscope analysis. In polarized Raman spectroscopy, the $B_{2g}$ mode shows the same polarization dependence regardless of the excitation wavelength or the sample thickness. On the other hand, the $A_g^1$ and $A_g^2$ modes show a peculiar polarization behavior that depends on the excitation wavelength and the sample thickness. The thickness dependence can be explained by considering the anisotropic interference effect due to birefringence and dichroism of the BP crystal, but the wavelength dependence cannot be explained. We propose a simple and fail-proof procedure to determine the orientation of a BP crystal by combining polarized Raman scattering with polarized optical microscopy.

**INTRODUCTION**

Two-dimensional layered materials including graphene, transition metal dichalcogenides, and black phosphorus (BP) have attracted much interest owing to their unique physical properties and superior electrical and/or mechanical characteristics. Among them, BP is drawing much attraction recently due to a high carrier mobility (300–1,000 V cm$^2$/s) and a high on/off ratio (~10$^5$) of BP field effect transistors.[1–5] Because BP is an anisotropic crystal, some of its physical properties such as mobility[2,3,6] and infrared light absorption[2,6–8] exhibit orientation dependence. Therefore, an easy method to determine the crystallographic orientation of BP crystals is needed in order to control orientation dependent properties of devices made of BP. Transmission electron microscopy (TEM) is a direct method to determine the orientation, but it is destructive and time consuming due to complex sample preparation procedures. Polarized Raman spectroscopy in combination with uniaxial strain is often used to determine the crystallographic orientation of 2-dimensional materials.[9–12] Because BP is anisotropic, one may expect that polarized Raman spectroscopy without strain can be used to determine the orientation of BP crystals. However, as we report here, the Raman modes in BP crystals show peculiar polarization behaviors that depend on the excitation wavelength and the sample thickness. Unless these dependences are carefully accounted for, one may make incorrect determination of the orientation. We show that in order to determine the orientation unambiguously, a short-wavelength excitation should be used or the interference effect should be explicitly accounted for. In addition, we propose an easy and fail-proof optical method to determine the crystallographic orientation of BP crystals by combining polarized optical microscopy and polarized Raman spectroscopy.

**RESULTS AND DISCUSSION**

**Electron and optical microscopy**

Among exfoliated BP samples, some flakes with long straight edges are found. In Fig. 1(a-d), such a flake is placed on a TEM grid and observed under a cross-polarized optical microscope in the reflection configuration. The incident light is polarized along the horizontal direction. When the long straight edge is parallel or orthogonal to the incident polarization, the sample appears dark, whereas it appears bright in intermediate directions. This effect will be discussed in more detail below. Figures 1(e) is the high-resolution

TEM (HR-TEM) image and the corresponding transmission electron diffraction (TED) pattern, respectively, near the long straight edge. It is established that the long straight edge is along the zigzag direction. This is consistent with a previous calculation that predicted that the ideal strength is much weaker for a tensile strain in the armchair direction, so that it is more likely that a crystal would cleave along the perpendicular direction, i.e., the zigzag direction.[13] Figure 1(f) shows the measured lateral atomic spacings which have similar values as the theoretical ones.[6]

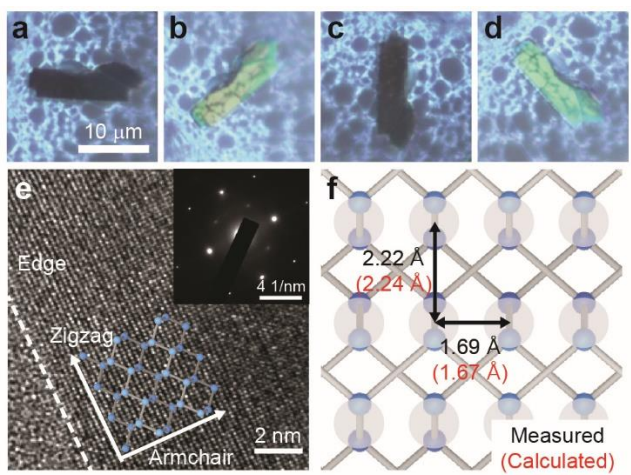

Fig. 1 (a-d) Polarized optical microscope images of a BP crystal on TEM grid. Incident light polarization (horizontal in the images) and the analyzer are orthogonal (cross polarization). (e) HR-TEM and TED (inset) images near the long straight edge of the sample. The dashed line is along the long straight edge. (f) Top view of the BP crystal structure. Two stacked layers are shown. Measured (black) and calculated (red) lateral atomic spacings are indicated.

The optical contrast of BP crystals was studied in more details by using cross-polarized optical microscopy on another sample of ~90 nm in thickness. In Fig. 2, the incident light polarization is along the horizontal direction and the analyzer is set along the vertical direction (cross polarization). The sample appears bright when the angle between the incident polarization and the long straight edge is 45 or 135 degrees, whereas it is dark when the angle is 0 or 90 degrees. In parallel polarization, the brightness of the sample does not change much because the surface reflection is rather strong regardless of the polarization direction. The behavior under cross polarization can be explained by birefringence. The crystal structure of bulk BP is orthorhombic (Cmce),[14,15] and the refractive indices along the three principal axes are different.[16] When linearly polarized light goes through a biaxial crystal with the polarization direction not parallel to one of the principal axes, the transmitted light becomes elliptically polarized, and thus optical contrast is observed even under cross polarization. The perpendicular component of the transmitted or reflected light is maximum when the angle between the incident polarization and the principal axis is 45° or 135°. From the comparison between these results with the TEM analysis of Fig. 1, we can determine that the long straight edge in this sample is also along the zigzag direction.

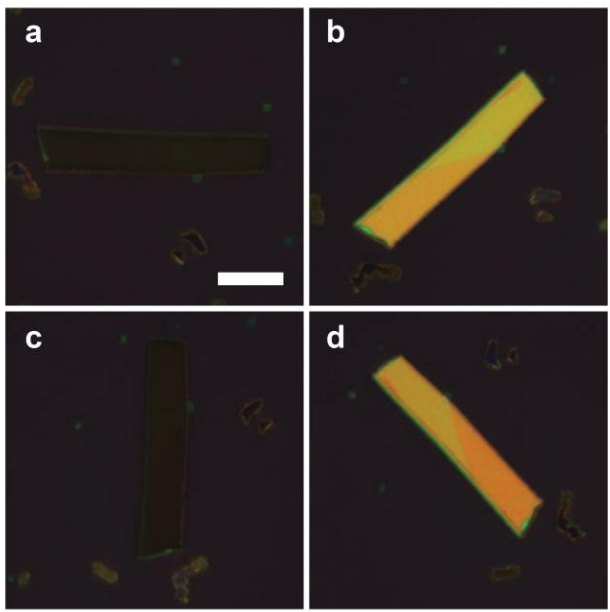

Fig. 2 Cross-polarized optical images of a BP crystal (thickness ~90 nm) on a SiO$_2$/Si substrate. The scale bar is 10 μm. The relative angle between the long straight edge (zigzag direction) and the incident polarization direction (horizontal) is (a) 0°, (b) 45°, (c) 90°, and (d) 135°.

Upon close inspection, we find that the sample appears slightly brighter for 45° than for 135° (Figure S1†). We compared several samples in both transmission and reflection configurations (Figure S2† and S3†) and found that the difference between 45° and 135° varies among samples. In transmission, the contrast difference between 45° and 135° is negligible. But in reflection, the contrast difference is often fairly significant (Figure S2†). This cannot be explained by simple (linear) birefringence because the two directions are symmetric in bulk BP. A possible effect of slight tilting of the sample with respect to the optical axis of the microscope was checked by intentionally tilting a sample. The effect was minimal. Circular dichroism or birefringence would explain the difference, but such chiral properties are not expected in BP due to symmetry. If symmetry is broken due to surface reconstruction,[17] it may become possible. We observed that the difference increased with time for a given sample, which implies that surface contamination may be responsible for the effect. Further investigation is needed to understand this phenomenon.

**Polarized Raman spectroscopy**

Polarized Raman measurements were carried out on the sample in Fig. 2. Figure 3(a) shows the crystal structure of BP. The zigzag direction is chosen to be along the *x* axis, and the armchair direction along the *z* axis. For polarized Raman measurements, the analyzer is set parallel to the incident polarization direction, and the spectra were measured as a function of the incident polarization with respect to the zigzag (*x*) direction.

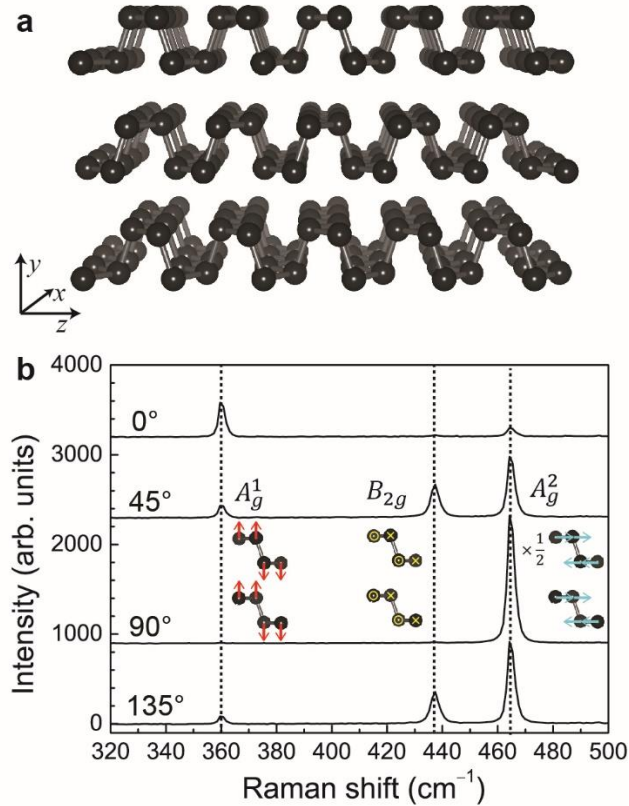

Fig. 3 (a) Crystal structure of BP. The principal axes are shown. (b) Polarized Raman spectra (parallel polarization) of BP measured with the 441.6-nm excitation. The angle between the incident polarization and the zigzag direction for each spectrum is indicated.

Figure 3(b) shows some representative polarized Raman spectra of BP measured with the 441.6-nm excitation. Three major peaks are observed. The peak at 360 cm$^{-1}$ is strongest at 0° and disappears at 90°. The peak at 437 cm$^{-1}$, on the other hand, disappears at 0° and 90°. The strongest peak at 464 cm$^{-1}$ is maximum at 90° and minimum at 0°, which is orthogonal to the one at 360 cm$^{-1}$. The polarization dependence of the peaks can be compared with group theoretical analysis to assign the modes as $A_g^1$, $B_{2g}$, and $A_g^2$, respectively. The vibrational modes are shown schematically in Fig. 3(b).

The intensity of polarized Raman signal is proportional to $|\hat{e}_i \cdot R \cdot \hat{e}_s|^2$, where $\hat{e}_i$ and $\hat{e}_s$ are the polarizations of the incident and scattered photons, respectively, and $R$ is the Raman tensor for a given mode. The Raman tensors of the Raman active modes in the backscattering geometry in complex form are: [14,15,18–20]

$$R(A_g) = \begin{pmatrix} |a|e^{i\phi_a} & 0 & 0 \\ 0 & |b|e^{i\phi_b} & 0 \\ 0 & 0 & |c|e^{i\phi_c} \end{pmatrix} \text{ and } R(B_{2g}) = \begin{pmatrix} 0 & 0 & |e|e^{i\phi_e} \\ 0 & 0 & 0 \\ |e|e^{i\phi_e} & 0 & 0 \end{pmatrix}. \quad (1)$$

It should be noted that the polarization dependence for parallel and cross polarizations cannot be fitted simultaneously without taking complex values for the tensor elements. The polarization vectors are given by $\hat{e}_i = \hat{e}_s = (\cos\theta, 0, \sin\theta)$ in backscattering geometry with parallel polarizations, where the angle $\theta$ is measured with respect to the zigzag direction. Then the Raman intensity is given by

$$I(A_g) \propto \left(|a|\cos^2\theta + |c|\cos\phi_{ca}\sin^2\theta\right)^2 + |c|^2 \sin^2\phi_{ca}\cos^4\theta \text{ and} \quad (2)$$

$$I(B_{2g}) \propto 4|e|^2 \cos^2\theta \sin^2\theta, \quad (3)$$

where $\phi_{ca} = \phi_c - \phi_a$. For cross polarization, $\hat{e}_i = (\cos\theta, 0, \sin\theta)$ and $\hat{e}_s = (\cos(\theta+90°), 0, \sin(\theta+90°)) = (-\sin\theta, 0, \cos\theta)$. The Raman intensity is given by

$$I(A_g) \propto \left(|a| - |c|\cos\phi_{ca}\right)^2 + |c|^2 \sin\phi_{ca}\cos^2\theta \text{ and} \quad (4)$$

$$I(B_{2g}) \propto |e|^2 \left(\cos^2\theta - \sin^2\theta\right)^2. \quad (5)$$

It is obvious that the zigzag (0°) and the armchair (90°) directions are clearly distinguished in the above analysis and the data in Fig. 3(b). These results are similar to previous reports. [2,19,21,22] However, as we shall see in the following, the polarization dependence of the modes is strongly influenced by the excitation wavelength and the thickness of the sample, and the crystallographic orientations cannot be determined unambiguously without taking this fact into account.

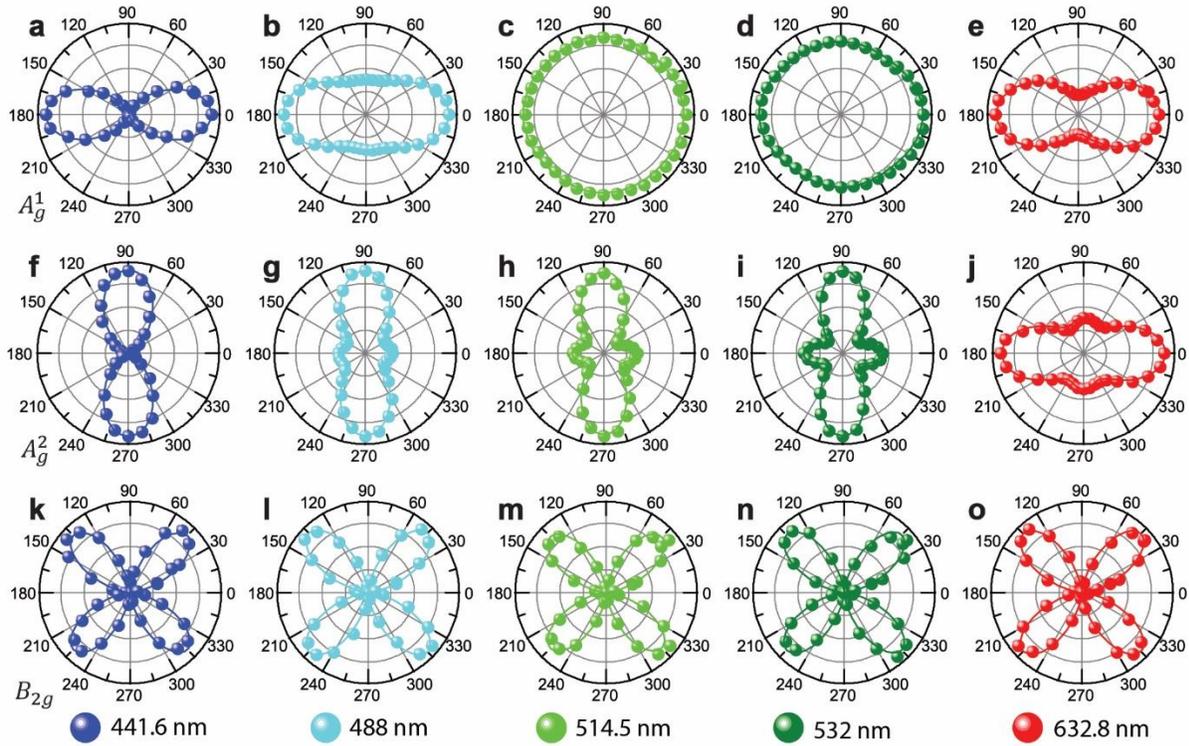

Fig. 4 Polarization dependence of Raman modes for different excitation wavelengths. Each row shows polarization dependence of $A_g^1$, $A_g^2$ and $B_{2g}$ modes, respectively, taken with excitation wavelengths of 441.6, 488, 514.5, 532, and 632.8 nm as indicated. The curves represent best fits to the calculated polarization dependence of the Raman intensities using Eqs. (2) and (3). The polarization behaviors of the $A_g$ modes exhibit strong dependence on the excitation wavelength, whereas the $B_{2g}$ mode does not show such dependence.

We repeated the polarized Raman measurements with different excitation lasers. Figure 4 summarizes the polarization dependence of the $A_g^1$, $A_g^2$, and $B_{2g}$ modes for excitation wavelengths of 441.6, 488, 514.5, 532, and 632.8 nm in parallel polarization. Similar data for cross polarization are shown in Figure S4†. The angles are measured with respect to the zigzag (*x*) direction. It is striking that the polarization dependence of the $A_g$ modes dramatically changes with the excitation wavelength. The $A_g^1$ mode, for example, shows a bow-tie shape dependence with the maximum at 0° for 441.6 nm. However, the same mode is almost isotropic for 514.5 or 532 nm. The $A_g^2$ mode is even more striking: the maximum occurs at 90° for all the excitation wavelengths except for 632.8 nm for which the maximum is at 0°. The $B_{2g}$ mode, on the other hand, has the same polarization dependence regardless of the excitation wavelength. We repeated the measurements with several samples with different thicknesses using 441.6-nm and 514.5-nm excitation wavelengths. The results are summarized in Figure 5 and the detail spectra are shown in Figure S5†. For different thickness, the polarization dependence for each mode is fairly similar for the 441.6-nm excitation, whereas it varies greatly with thickness for the 514.5-nm excitation.

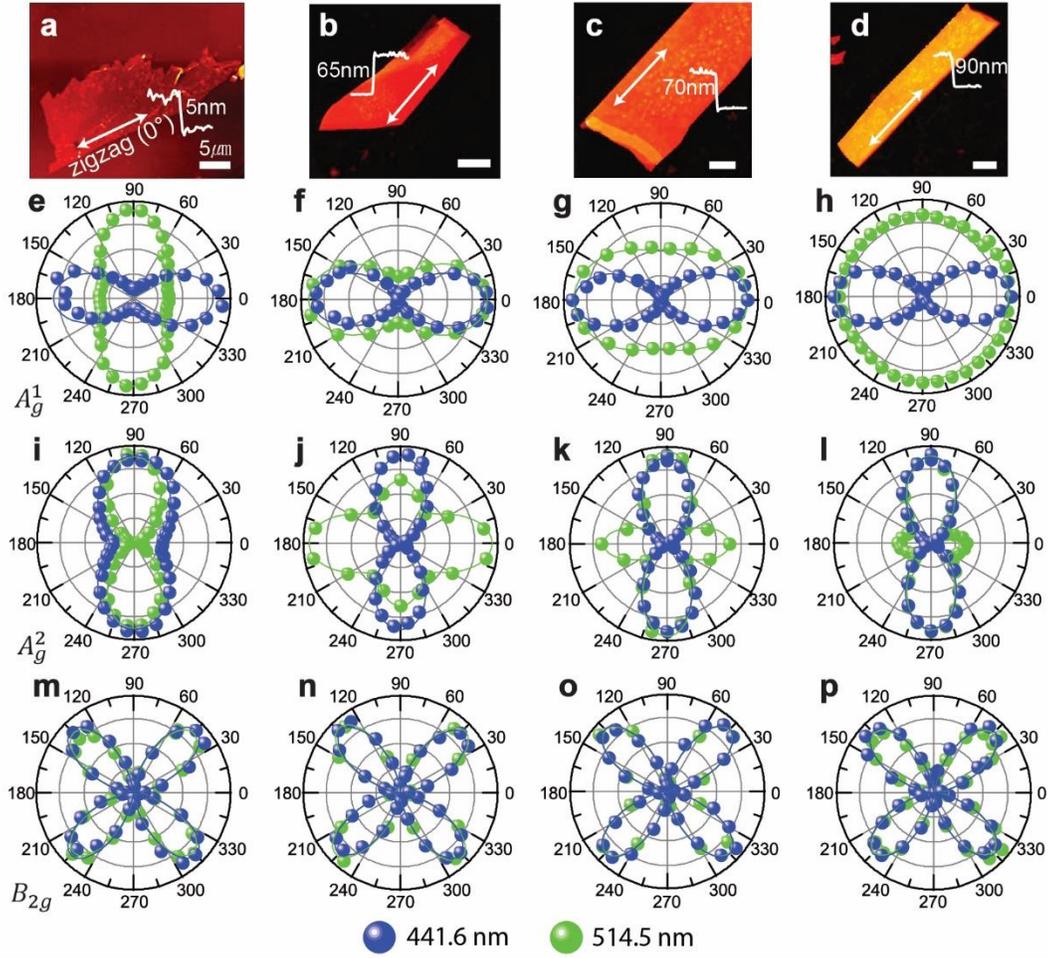

Fig. 5 Polarization dependence of Raman modes for different sample thickness. (a-d) are AFM images of 5-, 65-, 70- and 90-nm thick samples. The 5-nm sample is on a substrate with 100-nm $SiO_2$, and the others are on 300-nm $SiO_2$. Each row shows polarization dependence of $A_g^1$, $A_g^2$ and $B_{2g}$ modes, respectively, taken with excitation wavelengths of 441.6 and 514.5 nm as indicated. The curves represent best fits to the calculated polarization dependence of the Raman intensities using Eqs. (2) and (3). The polarization behaviors of the $A_g$ modes exhibit strong dependence on the thickness for the 514.5-nm excitation whereas no such dependence is seen for the 441.6-nm excitation.

The effect of optical interference[23] may explain this excitation wavelength and thickness dependence as both the wavelength and the sample thickness are the determining factors of interference. Because BP is an anisotropic material, it has both linear dichroism and birefringence, both of which contribute to polarization dependent interference. By using the reported refractive indices[24] we calculate the interference enhancement factors of the $A_g$ modes along the zigzag and armchair directions†. As shown in Fig. 6(a), the enhancement factors vary greatly with thickness and are dramatically different along the two principal axes. Figure 6(b) shows the ratio of the enhancement factors along the zigzag and the armchair directions. When the incident polarization is along the zigzag direction, the Raman intensity can be enhanced much more than in the armchair polarization case. Note that for the 441.6-nm excitation, the enhancement ratio does not vary much with thickness. This is because of the large imaginary part of the refractive index, $\kappa$, of BP at this wavelength. A large value of $\kappa$ suppresses multiple orders of interference and so the interference enhancement. Since $\kappa$ increases steeply between 488 nm and 441.6 nm (see Table S1†),[24] the effect of interference is strong for 488 nm and suppressed for 441.6 nm. Now one can plot the polarization dependence of the Raman modes with the interference effect corrected for. Figures 6(c-j) compare the effect of such corrections. After correction, the polarization dependence measured with the 514.5-nm excitation becomes fairly similar for the samples with different thickness. [Figs. 6(d, f)] Therefore, the sample thickness dependence can be explained as being due to anisotropic interference. However, for the different excitation wavelength, the $A_g^1$ and $A_g^2$ modes still shows qualitatively different polarization behaviors even after the correction for interference. Because of the anisotropy in the band structure, it is reasonable to expect that the Raman tensor elements vary with the excitation wavelength. In isotropic materials, the resonance effect reflects such an effect. In an anisotropic material such as BP, the excitation energy dependence of the tensor elements would result in variations in the polarization behaviors, and the $|a|/|c|$ ratio would change with the excitation wavelength. Such an effect is not observed in isotropic 2-dimensional materials such as graphene or $MoS_2$.[25,26] On the other hand, because $R(B_{2g})$ has only one element $e$, the functional form of the polarization dependence of the $B_{2g}$ mode will not change even if the magnitude of $e$ changes. This would explain the observed difference between the $A_g$ modes and the $B_{2g}$ mode. The polarization dependence of the $A_g$ modes for each excitation wavelength in Figs. 6(h, j) is fitted to the calculated Raman intensity using Eqs. (2) and (3), and the obtained $|a|/|c|$ ratios and $\phi_{ca}$ values are summarized in Table 1.

Table 1. Excitation wavelength dependence of *a/c* ratio. Ratios of the Raman tensor components *a* and *c* were obtained from fitting the data in Fig. 4 to Eqs. (2) and (3).

| Wavelength (nm) | $A_g^1$ | | $A_g^2$ | |
|---|---|---|---|---|
| | a/c | phase (°) | a/c | phase (°) |
| 441.6 | 4.01 | 0 | 0.24 | 179 |
| 488 | 1.20 | 48 | 0.47 | 97 |
| 514.5 | 0.76 | 34 | 0.50 | 111 |
| 532 | 0.75 | 43 | 0.52 | 120 |
| 632.8 | 0.98 | 47 | 0.79 | 91 |

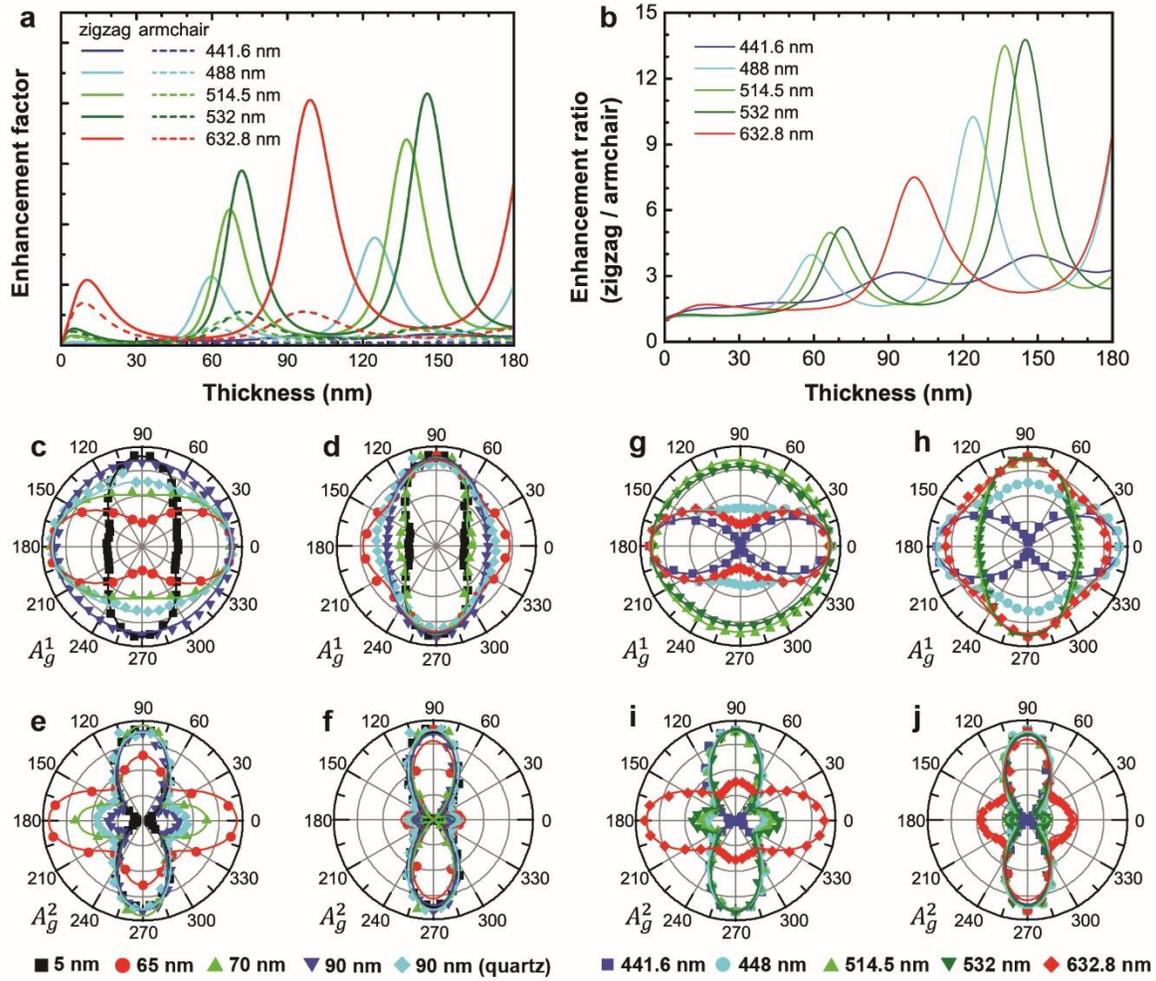

Fig. 6 (a) Calculated interference enhancement factor as a function of the sample thickness for zigzag (*x*) and armchair (*z*) directions for different excitation wavelengths for BP on 300 nm SiO$_2$ on Si substrate. (b) Enhancement ratio obtained by dividing the enhancement factor for the zigzag direction with that for the armchair direction. (c-f) Polarization dependence of $A_g$ modes for different thicknesses measured with 514.5-nm excitation, (c, e) before and (d, f) after the correction for interference. The data include a 5-nm thick sample on 100-nm SiO$_2$ on Si and a 90-nm thick sample on quartz. (g-j) Polarization dependence of $A_g$ modes for different excitation wavelengths from a 90-nm thick sample on 300-nm SiO$_2$ on Si, (g, i) before and (h, j) after the correction for interference.

In order to visualize the thickness dependence, we performed polarized wide-field Raman imaging measurements[27] on a sample with several BP flakes with different thicknesses using the 514.5-nm excitation. The wide-field Raman images for each mode as the polarization (indicated by the arrow) of the incident laser is rotated are shown in Movie S1†. Figure 7(a) is an optical image of the sample. BP flakes with different thicknesses appear in different colors. Figure 7(b) represents the intensity ratio of the $A_g^2$ mode intensity taken with the excitation polarization in the horizontal direction to that in the vertical direction. For this measurement, two wide-field Raman images were taken from the same sample area successively with the two polarization directions, and then the ratio image was obtained by processing the images. Figure 7(b) clearly shows that even in the same flake, areas with different thickness have different ratios. Therefore, relying on the polarization behavior of the $A_g$ modes alone may lead to incorrect determination of the crystallographic orientation if the excitation wavelength is 514.5 nm or longer.

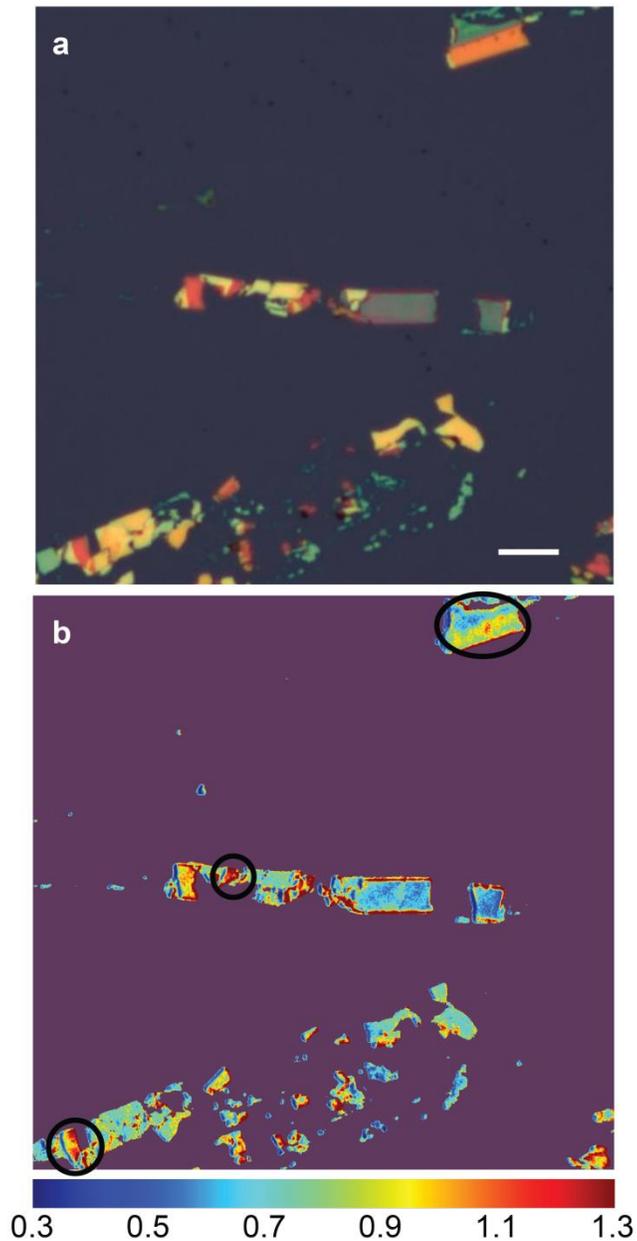

Fig. 7 (a) Optical image of a sample with several BP flakes with different thicknesses. The scale bar is 20 μm. (b) Polarized Raman intensity ratio image. The Raman intensity of the $A_g^2$ mode taken with the excitation polarization in the horizontal direction is divided by that in the vertical direction. The intensity ratio varies greatly even inside a given flake, depending on the thickness (circled areas).

## CONCLUSIONS

Based on our results, we propose the following procedure to determine the crystallographic orientation of BP flakes. First, the sample should be inspected by an optical microscope in cross polarization. By rotating the sample until the sample appears dark, one can determine that the zigzag direction (*x*) is either parallel or perpendicular to the incident polarization or the analyzer direction. At this point, the principal axes of the crystal (*x* or *z*) are found, but one cannot know which of the two directions the zigzag direction is. Then, two Raman spectra in parallel polarization need to be measured, with the excitation polarization along each of the principal axes. One should use a short wavelength laser (441.6 nm in our case). The direction in which the $A_g^1$ ($A_g^2$) mode is stronger (weaker) is the zigzag direction. One can indeed skip the first step with an optical microscope if a full polar plot of one of the $A_g$ modes is measured with a short wavelength laser. However, because many more Raman spectra should be taken, it would be much more time consuming.

## EXPERIMENTAL METHODS

### Sample preparations

The samples were prepared directly on $SiO_2$/Si substrates by mechanical exfoliation from BP flakes (Smart elements). For HR-TEM measurements, exfoliated BP flakes were transferred from the Si/$SiO_2$ wafer onto a TEM grid using the wet transfer method.[28] A wafer with BP flakes was coated with poly (methyl methacrylate) (PMMA) at 4000 RPM and separated in a buffered oxide etchant (BOE) solution by etching the $SiO_2$ layer. To prevent contaminants, it was washed in DI water several times. Then, BP with PMMA support was transferred onto a TEM grid (Lacey Formvar/Carbon Coated Copper Grid, SPI Supplies). To remove the PMMA support, the grid sample was put into acetone for 15 min and annealed in a flow of Ar gas at 250℃ for 9h. The optical measurements were carried out with BP samples on $SiO_2$/Si substrates. The thickness was determined by using an atomic force microscope (NT-MDT NTEGRA Spectra). The samples were kept in an optical vacuum chamber to keep them from contamination in atmosphere.[29,30]

**Polarized optical microscopy**

A commercial optical microscope (Nikon LV100) was used to obtain the optical images. The incident polarization direction was fixed and the sample was rotated by using a rotation stage. The analyzer was fixed to cross polarization configuration (perpendicular to incident polarization).

**HR-TEM measurements**

HR-TEM analysis was carried out using a Cs-corrected high-resolution transmission electron microscope (FEI Titan) operated in the range of 80 to 200 kV.

**Polarized Raman measurements**

We used 5 different excitation sources: the 441.6-nm (2.81 eV) line of a He-Cd laser, the 488 and 514.5-nm (2.54 and 2.41 eV) lines of an Ar ion laser, the 532-nm (2.33 eV) line of a diode-pumped-solid-state laser, and the 632.8-nm (1.96 eV) line of a He-Ne laser. To prevent degradation of the sample,[29] all the measurements were performed while the sample was kept in an optical vacuum chamber (Oxford Microstat He2). The laser beam was focused onto the sample by a 40× long-working distance microscope objective lens (0.6 N.A.), and the scattered light was collected and collimated by the same objective. The Raman signal was dispersed with a Jobin-Yvon Horiba TRIAX 550 spectrometer (1800 grooves/mm) and detected with a liquid-nitrogen-cooled back-illuminated charge-coupled-device (CCD) detector. The laser power was kept below 0.5 mW. An achromatic half-wave plate was used to rotate the polarization of the linearly polarized laser beam to the desired direction. For parallel (perpendicular) polarization, the analyzer angle was set such that photons with polarization parallel (perpendicular) to the incident polarization pass through. Another achromatic half-wave plate was placed in front of the spectrometer to keep the polarization direction of the signal entering the spectrometer constant with respect to the groove direction of the grating.[25,31]

To obtain polarized wide field Raman images, a liquid-crystal tunable bandpass filter (Varyspec VISR) and an electron multiplying charge-coupled-device (EMCCD, Andor iXon3 888) were used. For illumination, the incident laser beam was passed through a beam shaper (piShaper 6_6) to make the power density uniform across the illuminated area of ~200 μm in diameter. The total laser power was 100 mW, and the spectral resolution was ~ 10 cm$^{-1}$.

**Calculation of enhancement factors by interference effects**

We used the model that Yoon *et al.*[23] used to calculate the intensity enhancement of Raman scattered light by the interference effect. The laser beam is absorbed during the multiple reflections. The net absorption term at position $x$ [ $F_{ab}(x)$ ] can be expressed as

$$F_{ab}(x) = t_{01} \frac{(1+r_{12}r_{23}e^{-2i\beta_2^{ex}})e^{-i\beta_x^{ex}} + (r_{12}+r_{23}e^{-i\beta_2^{ex}})e^{-i(2\beta_1^{ex}-\beta_x^{ex})}}{1+r_{12}r_{23}e^{-2i\beta_2^{ex}} + (r_{12}+r_{23}e^{-2i\beta_2^{ex}})r_1 e^{-2i\beta_1^{ex}}},$$

where $t_{ij} = 2n_i/(\tilde{n}_i + \tilde{n}_j)$ and $r_{ij} = (\tilde{n}_i - \tilde{n}_j)/(\tilde{n}_i + \tilde{n}_j)$ are the Fresnel coefficients at the interfaces of the *i*-th and the *j*-th layer; the indices are assigned as air (0), BP (1), SiO$_2$ (2), and Si (3). $\tilde{n}_i$ is the complex refractive index for the *i*-th layer. The phase terms $\beta_x^{ex} = 2\pi x \tilde{n}_1/\lambda_{ex}$ and $\beta_i^{ex} = 2\pi d_i \tilde{n}_i/\lambda_{ex}$ are included, where $d_i$ is the thickness of the *i*-th layer and $\lambda_{ex}$ the excitation wavelength.

The net scattering term [$F_{sc}(x)$] by multiple reflection of the Raman signal generated at position *x* can be written as

$$F_{sc}(x) = t_{10} \frac{(1+r_{12}r_{23}e^{-2i\beta_2^{sc}})e^{-i\beta_x^{sc}} + (r_{12}+r_{23}e^{-i\beta_2^{sc}})e^{-i(2\beta_1^{sc}-\beta_x^{sc})}}{1+r_{12}r_{23}e^{-2i\beta_2^{sc}} + (r_{12}+r_{23}e^{-2i\beta_2^{sc}})r_1 e^{-2i\beta_1^{sc}}},$$

where $\beta_x^{sc} = 2\pi x \tilde{n}_1/\lambda_{sc}$ and $\beta_i^{sc} = 2\pi d_i \tilde{n}_i/\lambda_{sc}$ are the phase terms with $\lambda_{sc}$ being the wavelength of the Raman signal. By considering the above two terms, the total enhancement factor (*F*) is given by

$$F = N \int_0^{d_1} |F_{ab}(x) F_{sc}(x)|^2 dx$$

where *N* is the normalization factor. We used the reported refractive indices by Asahina *et al.*[24] Since the refractive indices are different for zigzag and armchair directions of black phosphorus, the enhancement factors are calculated separately. The used values are summarized in Table S1†.

ACKNOWLEDGMENT

This work was supported by the National Research Foundation (NRF) grants funded by the Korean government (MSIP) (Nos. 2011-0013461, 2011-0017605 and 2015R1A2A2A01006992) and by a grant (No. 2011-0031630) from the Center for Advanced Soft Electronics under the Global Frontier Research Program of MSIP.

NOTES AND REFERENCES

† Electronic Supplementary Information (ESI) available. See DOI: 10.1039/b000000x

‡ J.K., J.-U.L., and J.L. contributed equally.

**Table of Contents**

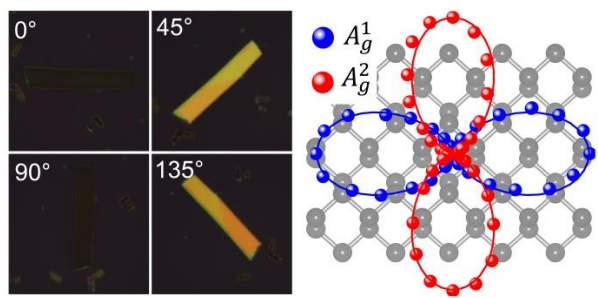

Crystallographic direction of black phosphorus can be determined from polarized optical microscopy and polarized Raman spectroscopy

# Electronic Supplementary Information (ESI)

# Anomalous polarization dependence of Raman scattering and crystallographic orientation of black phosphorus

*Jungcheol Kim,* [a, ‡] *Jae-Ung Lee,* [a, ‡] *Jinhwan Lee,* [b, ‡] *Hyo Ju Park,*[c] *Zonghoon Lee,*[c] *Changgu Lee* [b, d ,*] *and Hyeonsik Cheong* [a,*]

[a] Department of Physics, Sogang University, Seoul 121-742, Korea

[b] Department of Mechanical Engineering and Center for Human Interface Nano Technology (HINT), Sungkyunkwan University, Suwon, 440-746, Korea

[c] School of Materials Science and Engineering, Ulsan National Institute of Science and Technology (UNIST), Ulsan 689-798, Korea

[d] SKKU Advanced Institute of Nanotechnology (SAINT), Sungkyunkwan University, Suwon, 440-746, Korea

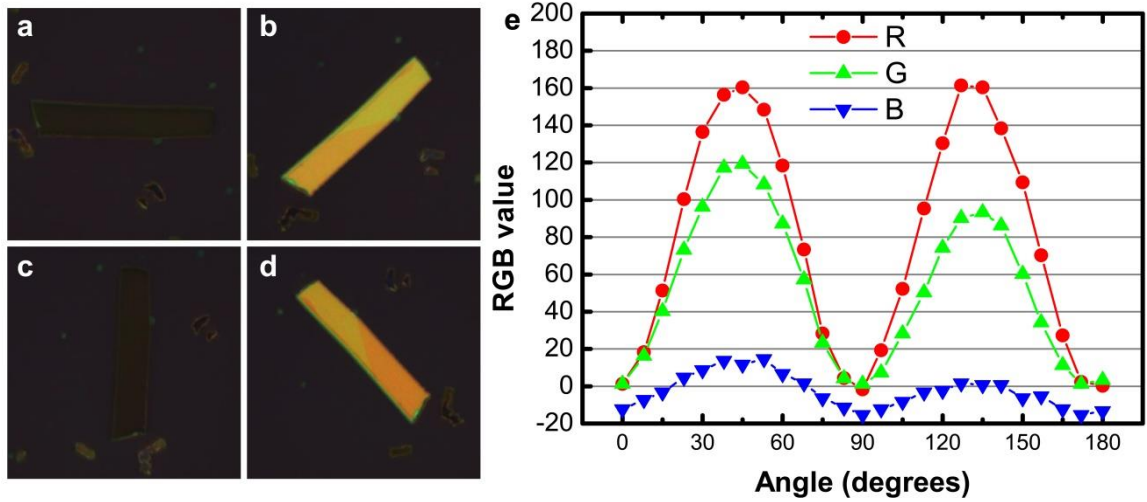

Fig. S1 Orientation dependence of color of a BP crystal on a SiO$_2$/Si substrate in reflection mode. (a-d) Optical images taken with reflected light in cross polarization. (e) The RGB values of the sample image with respect to that of the substrate as a function of the angle between the incident polarization and the long straight edge. There is a small difference between 45° and 135°.

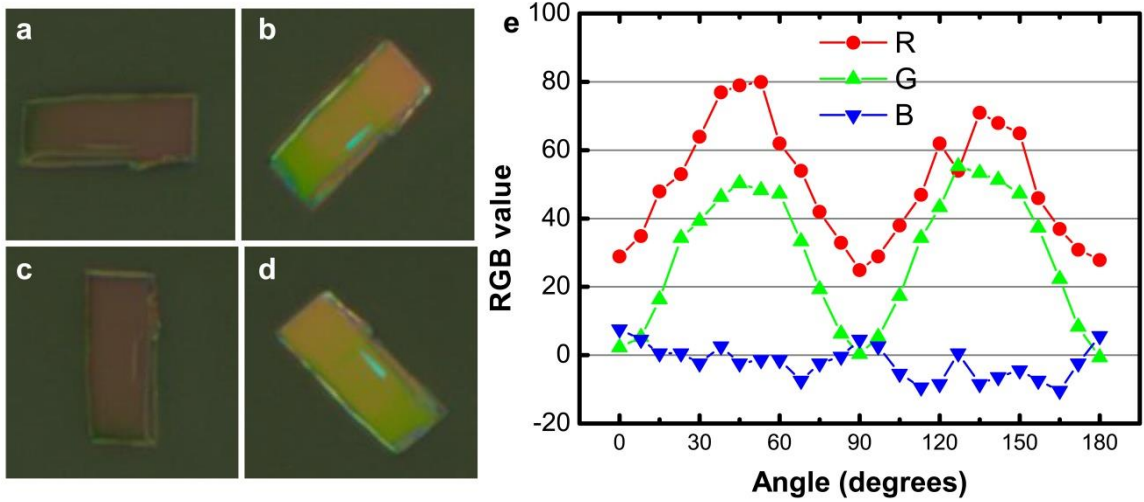

Fig. S2 Orientation dependence of color of a BP crystal on slide glass in reflection mode. (a-d) Optical images taken with reflected light in cross polarization. (e) The RGB values of the sample image with respect to that of the substrate as a function of the angle between the incident polarization and the long straight edge. There is a small difference between 45° and 135°.

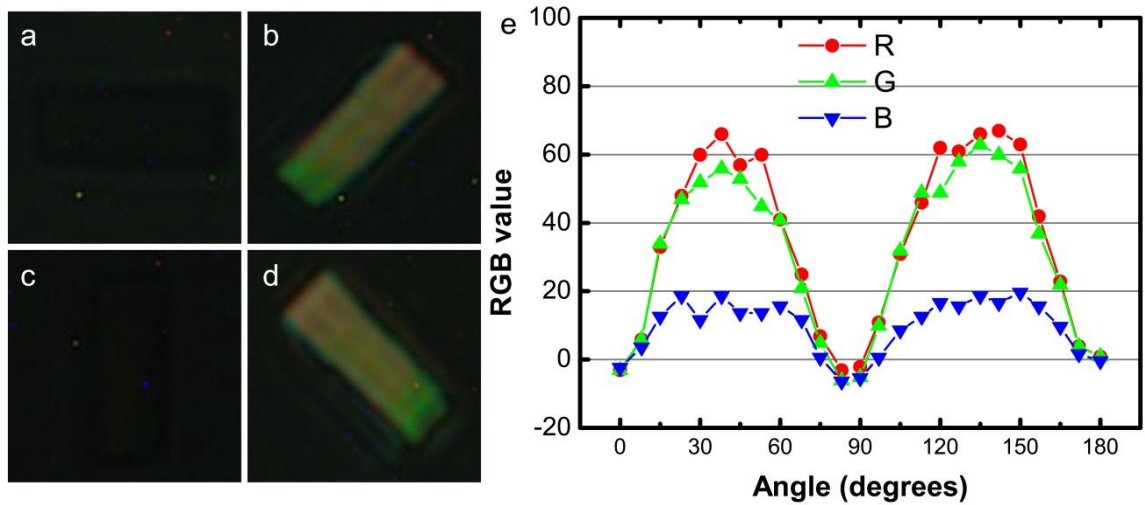

Fig. S3 Orientation dependence of color of a BP crystal on slide glass in transmission mode. (a-d) Optical images taken with transmitted light in cross polarization. (e) The RGB values of the sample image with respect to that of the substrate as a function of the angle between the incident polarization and the long straight edge. There is no appreciable difference between 45° and 135°.

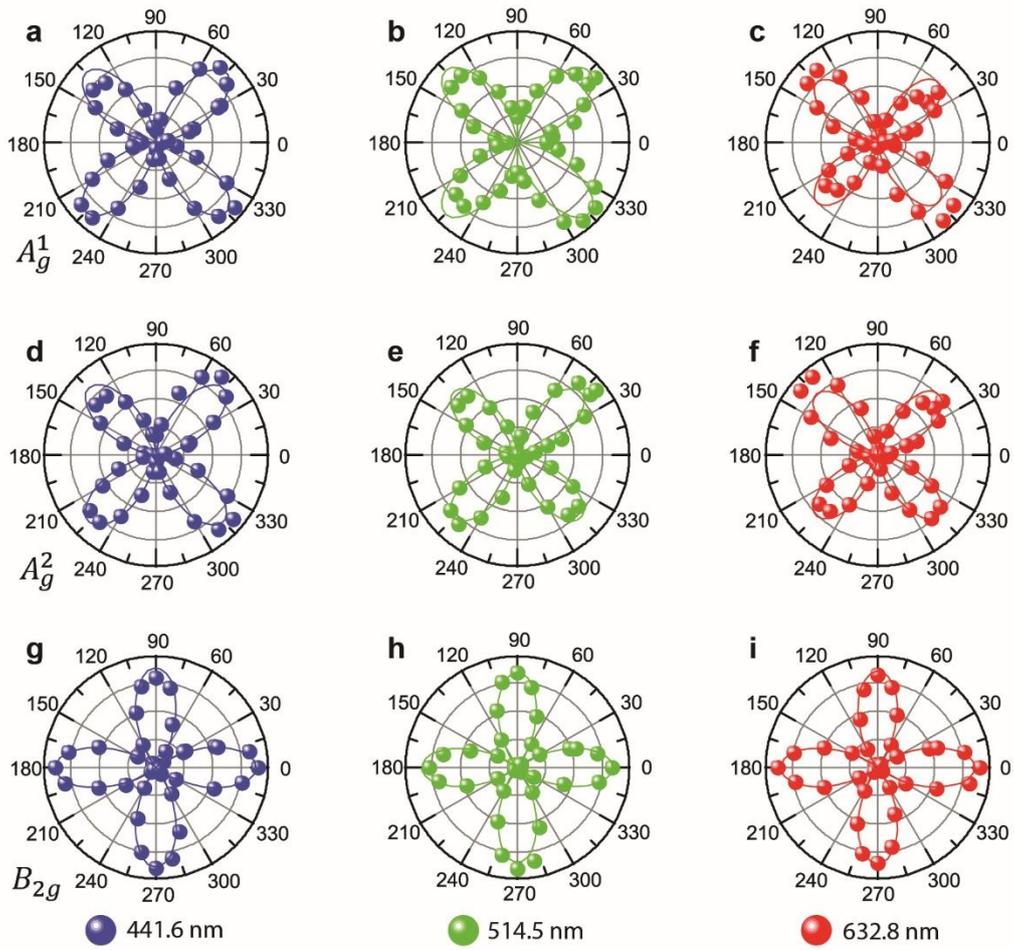

Fig. S4 Polarization dependence of Raman modes in cross polarization. Each row shows polarization dependence of $A_g^1$, $A_g^2$ and $B_{2g}$ modes, respectively, taken with excitation wavelengths of 441.6, 514.5, 632.8 nm as indicated.

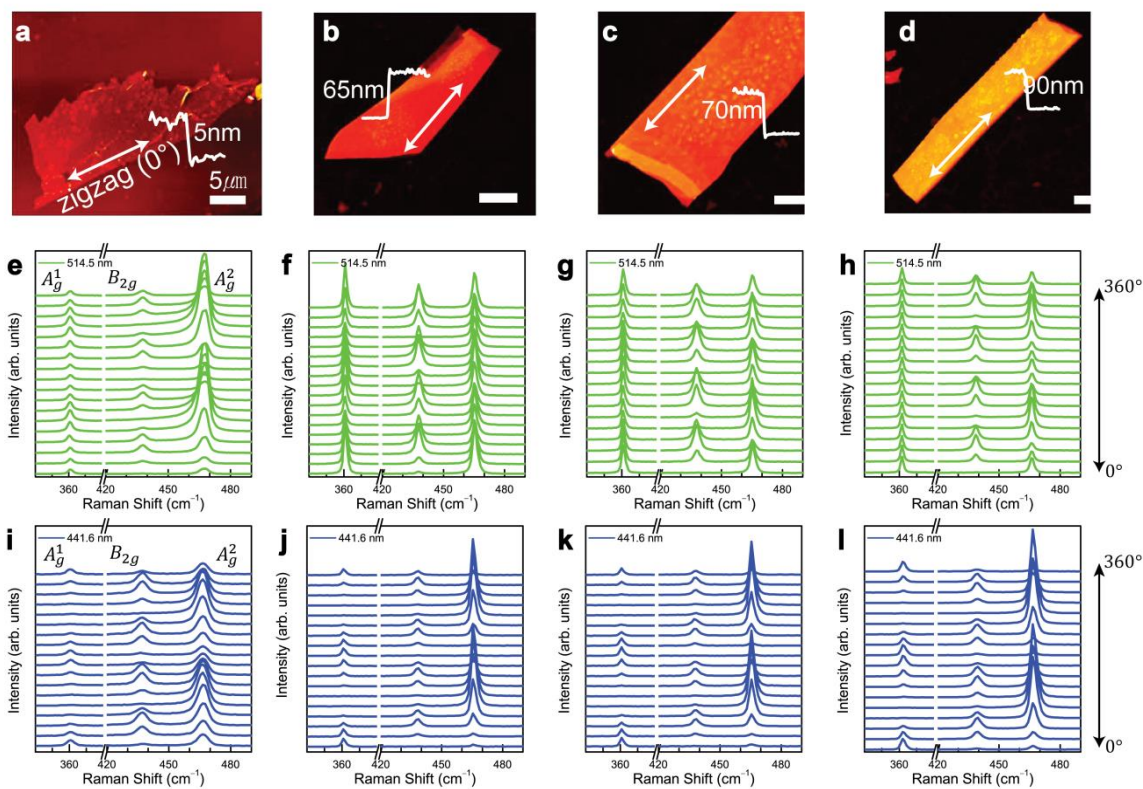

Fig. S5 Raman spectra corresponding to data in Fig. 5(e-p). The spectra are shown in 20-degree increments. (e-h) and (i-l) are taken with the excitation wavelengths of 514.5 and 441.6 nm, respectively.

Table S1 Refractive indices used in the calculation.

| Wavelength (nm) | BP (zigzag)[1] | BP (armchair)[1] | SiO$_2$[2] | Si[3] |
|---|---|---|---|---|
| 441.6 | 4.10-0.21i | 3.92-0.94i | 1.47 | 4.79-0.10i |
| 449.1 | 4.06-0.15i | 3.88-0.65i | 1.47 | 4.70-0.090i |
| 450.6 | 4.05-0.14i | 3.87-0.64i | 1.47 | 4.68-0.087i |
| 488.0 | 3.76-0.064i | 3.66-0.45i | 1.46 | 4.36-0.035i |
| 497.1 | 3.72-0.060i | 3.64-0.43i | 1.46 | 4.30-0.027i |
| 499.1 | 3.71-0.060i | 3.64-0.43i | 1.46 | 4.29-0.025i |
| 514.5 | 3.67-0.050i | 3.60-0.40i | 1.46 | 4.21-0.016i |
| 524.6 | 3.65-0.050i | 3.58-0.37i | 1.46 | 4.17-0.012i |
| 526.8 | 3.64-0.050i | 3.57-0.37i | 1.46 | 4.15-0.011i |
| 532.0 | 3.62-0.050i | 3.57-0.37i | 1.46 | 4.21-0.010i |
| 542.8 | 3.61-0.050i | 3.54-0.36i | 1.46 | 4.10-0.0077i |
| 545.1 | 3.60-0.050i | 3.53-0.37i | 1.46 | 4.22-0.0075i |
| 632.8 | 3.46-0.044i | 3.50-0.39i | 1.46 | 4.14-0.0010i |
| 648.1 | 3.43-0.040i | 3.48-0.36i | 1.46 | 4.16-0.0015i |
| 651.4 | 3.43-0.040i | 3.48-0.37i | 1.46 | 4.09-0.0016i |